\newcommand{\be}{\begin{equation}}
\newcommand{\ee}{\end{equation}}
\newcommand{\bea}{\begin{eqnarray}}
\newcommand{\eea}{\end{eqnarray}}

\documentstyle[aps,prd,tighten,preprint,floats]{revtex}
\begin{document}
\textwidth 6.2in
\textheight 8.7in
\thispagestyle{empty}
\topmargin -0.20in
\oddsidemargin 0.25in
\evensidemargin 0.25in
\baselineskip=20pt
\renewcommand{\thesection}{\Roman{section}}

\title{Towards a Gravitational Analog to $S$-duality in
Non-abelian Gauge Theories}
\author{H. Garc\'{\i}a-Compe\'an$^{a}$\thanks{Present Address:
{\it School of Natural Sciences,
Institute for Advanced Study, Olden Lane, Princeton NJ 08540 USA}. E-mail:
compean@sns.ias.edu}, O. Obreg\'on$^{b}$\thanks{E-mail:
octavio@ifug3.ugto.mx}, J.F. Pleba\'nski$^a$\thanks{E-mail:
pleban@fis.cinvestav.mx} and C. Ram\'{\i}rez$^c$\thanks{E-mail:
cramirez@fcfm.buap.mx}\\
$^{a}$ {\it Departamento de F\'{\i}sica, Centro de Investigaci\'on y de
Estudios
Avanzados del IPN\\
 Apdo. Postal 14-740, 07000, M\'exico D.F., M\'exico}\\
$^b$ {\it Instituto de F\'{\i}sica de la Universidad de Guanajuato\\
 P.O. Box E-143, 37150, Le\'on Gto., M\'exico}\\
$^c$ {\it  Facultad de Ciencias F\'{\i}sico Matem\'aticas\\
Universidad Aut\'onoma de Puebla\\
A.P. 1364, Puebla 72000, M\'exico}}

\maketitle

\begin{abstract}
For non-abelian non-supersymmetric gauge theories, generic  dual
theories have been constructed.  In these theories the couplings
appear inverted.  However,  they do not possess a Yang-Mills
structure but rather are a kind  of  non-linear
sigma model.  It is shown that for a topological gravitational
model an analog to this duality exists.
\end{abstract}

PACS number(s):  04.20.Cv, 04.50.+h, 11.15.-q, 11.30.Ly

\newpage
\renewcommand{\thesection}{\arabic{section}}
\renewcommand{\theequation}{\thesection.\arabic{equation}}

\section{Introduction}
\setcounter{equation}{0}

Recently $S$-duality in supersymmetric Yang-Mills theories and in
superstring theories, has become a most powerful non-perturbative
technique to describe and to compute strong coupling dynamics in these
theories.

It is well known that even perturbatively, superstring theories describe a
consistent (ultraviolet finite and unitary) theory of quantum gravity.
However,  very important gravitational issues remain behind the
non-perturbative sector.  For instance, the structure of the microstates
used in the computation of the entropy of extremal and near extremal
black holes \cite{Strominger}.

Roughly speaking, for the effective low energy action of string theories,
 $S$-duality symmetry is realized at the level of the axion and dilaton
moduli. The gravitational sector appears dynamical with respect to this
symmetry. In fact,  it is well known that gravitational corrections are
required in order to test string duality \cite{Vafa}, and to check some
consistency conditions in $M$-theory \cite{Witten}.

For the heterotic string theory in ten dimensions, toroidal
compactification, to four dimensions on the six-torus, gives for its low
energy limit, ${\cal N}=4$ super Yang-Mills theory on ${\bf R}^4$. Thus
$S$-duality in the four dimensional effective theory comes as a
consequence of superstring dualities in ten dimensions. The
four-dimensional effective theory is decoupled from gravitational effects,
and gravity enters only as an spectator (not dynamical). From this theory
a twisted topological field theory can be constructed on a curved
four-manifold,  which it is shown to be $S$-dual (according to the
Montonen-Olive conjecture), by using different formulas of four-manifolds
well known by the topologists \cite{Vafaone}.

On the side of non-supersymmetric gauge theories in four dimensions, the
subject has been explored recently in the abelian as well as in the
non-abelian cases  \cite{Wittenone,Verlinde,Barbon,Lozano,Ganor,Mohammedi},
see also \cite{Halpern,Fradkin,Rocek}. In the
abelian case, one considers $CP$ non-conserving Maxwell theory on a
curved compact four-manifold $X$ with Euclidean signature or, in other
words, U(1) gauge theory with a $\theta$ vacuum coupled to four-dimensional
gravity. The manifold $X$ is basically described by its associated
classical topological invariants: the Euler characteristic $\chi(X)  = { 1
\over 16 \pi^2}\int_X {\rm tr} R \wedge  \tilde R$ and the signature
$\sigma(X) = - {1 \over 24 \pi^2}\int_X {\rm tr} R\wedge R$.

In the Maxwell theory, the partition function $Z(\tau)$ transforms as a
modular form under a finite index subgroup $\Gamma_0(2)$
of SL$(2,{\bf Z})$ \cite{Wittenone,Verlinde}, $Z(-1/\tau) =
\tau^u \bar{\tau}^v Z(\tau)$, with the modular weight $(u,v) =
({1 \over 4}(\chi + \sigma), {1\over 4}(\chi - \sigma))$. In the above
formula $ \tau = {\theta \over 2 \pi} + {4 \pi i \over g^2}$,  where $g$
is the U(1) electromagnetic coupling constant and $\theta$ is the usual
theta angle.

Witten has shown \cite{Wittenone} that, at the quantum level, in order to
cancel the modular anomaly in abelian theories, one has to choose certain
holomorphic couplings $B(\tau)$ and $C(\tau)$ in  the
topological gravitational (non-dynamical) sector, through the
 action

\be
 I^{TOP} = \int_X \bigg( B(\tau) {\rm tr} R \wedge \tilde R + C(\tau)
{\rm tr} R \wedge R \bigg),
\label{1}
\ee
i.e., which is proportional to the appropriate sum of the Euler characteristic
$\chi(X)$ and the signature $\sigma(X)$. These couplings must satisfy the
condition that $exp(B(\tau))$ and $exp(C(\tau))$ should be modular forms
and their associate weights are chosen just to cancel the anomaly
\cite{Wittenone}. Therefore,  we must  consider, for instance, $CP$
non-conserving Maxwell theory with action $I_{M, \theta}$ coupled to
$I_{TOP}$ such that

\be
 I = I_{M, \theta} + \int_X \bigg( B(\tau) {\rm tr} R \wedge \tilde R +
C(\tau)  {\rm tr} R \wedge R \bigg).
\label{2}
\ee
$I_{M,\theta}$ consists in a dynamical plus a $\theta$-topological term and
$I^{TOP}$ works as a counterterm which cancels the modular anomaly.
Furthermore, if one considers the path integration over all phase space of
the theory, then it can be shown \cite{Lozano} that the path integral
phase space measure $\int {\cal D}A_{\alpha} {\cal D} \pi^{\alpha},$ where
$A_{\alpha}$ represents the abelian connection and $\pi^{\alpha}$ its
conjugate momenta,  codifies the modular anomaly in such a way that the
phase space partition function $Z_{ps}$ is in fact a modular invariant,
$Z_{ps}(\tau)  = Z_{ps}(- 1/\tau)$.

In the case of Yang-Mills theories, the lack of a generalized Poincar\'{e}
lemma means that there is no dual theory in the same sense as for abelian
theories \cite{chan}.  For  generic Yang-Mills theories,
one can follow the same procedure as in the abelian case
\cite{Lozano,Ganor,Mohammedi}. Usually one
 constructs  an {\it intermediate Lagrangian} from which one recuperates
the original Lagrangian and its dual, as different limits (for a recent
review, see \cite{Quevedo})

\be
L=- \alpha G_{\mu\nu}^a G^{\mu\nu}_ a+G^{\mu\nu}_a
F_{\mu\nu}^a(A),  \label{parent}
\ee
$\alpha$ is the constant coupling, $F_{\mu\nu}^a$ is a Lie
algebra-valued tensor field and $G^a_{\mu \nu}$ is a Lie algebra-valued
Lagrange multiplier tensor field. Obviously, if the variables $G$ are
integrated out, we get the usual Yang-Mills  Lagrangian
$L= {1\over{4 \alpha}} F_{\mu\nu}^a F^{ a \mu\nu}$,  where
  $F^a_{ \mu \nu} = \partial_{\mu} A^a_{\nu} - \partial_{\nu}
A^a_{\mu} + f^a_{bc}A^b_{\mu}A^c_{\nu}$ and $f^a_{bc}$ are the structure
constants of the Lie algebra of the gauge group.

Thus, the Euclidean partition function of (\ref{parent}), after partial
integration, will be given by

\begin{equation}
Z=\int {\cal D}G exp \big( -\int \alpha G_{\mu\nu}^a G^{\mu\nu}_
a dx \big) \int {\cal D} A exp \bigg[ \int(2\partial_\mu G^{\mu\nu}_a
A_\nu^a-M^{\mu\nu}_{ab} A_\mu^a A_\nu^b) dx \bigg],
\label{z}
\end{equation}
$ M^{\mu\nu}_{ab}=f^c_{ab}G^{\mu\nu}_ c$ is the adjoint transformed
of $G$.  Furthermore, this partition function can be rewritten as

\begin{equation}
\begin{array}{ll}
Z=&\int {\cal D}G
exp \big[ - \int (\alpha G_{\mu\nu}^a G^{\mu\nu}_ a -
{M^{-1}}_{\mu\nu}^{ab}\partial_\rho G^{\nu\rho}_b\partial_\tau G^{\mu\tau
}_a )dx \big] \int {\cal D} Aexp ( \int M^{\mu\nu}_{ab}A_\mu^aA_\nu^b dx ) \\
=&\sqrt{\pi} \int {\cal D}G \sqrt{det(M^{-1})}exp \big[ - \int (\alpha
G_{\mu\nu}^a G^{\mu\nu}_ a - {M^{-1}}_{\mu\nu}^{ab}\partial_\rho
G^{\nu\rho}_b\partial_\tau G^{\mu\tau }_a )dx \big],
\label{z1}
\end{array}
\end{equation}
which represents in some sense the dual of the starting Yang-Mills
theory.
Obviously, this is a complicated ``massive" non-linear sigma model
\cite{Freed}  which
could not be well defined, depending on if the ``metric'' $M$ is singular
or not. Gannor and Sonnenschein \cite{Ganor} show how to regain a
Yang-Mills theory from this, in such a way that it apparently leads to the
dual theory. Following them, let us define
$\bar{A}_\mu^a = - ({M^{-1}})_{\mu\nu}^{ab}\partial_\rho G^{\nu\rho}_b$,
that is, $G$ satisfies the equations of motion
$\partial_\nu G^{\nu\mu}_b+M^{\mu\nu}_{ab}
\bar{A}_\nu^a=0$.
In abelian theories, the Poincar\'e lemma gives the solution to this
equation in terms of a vector potential for the dual of $G$. However, for
non-abelian theories, although $F_{\mu\nu}^a(\bar{A}) = \partial_\mu
\bar{A}^a_\nu - \partial_\nu \bar{A}^a_\mu + f^a_{bc} \bar{A}^b_\mu
\bar{A}^c_\nu$
 is a solution for $G^a_{\mu\nu}$, it is not the most
general solution.

Nevertheless, it can be easily seen that the second term in the
exponential of (\ref{z1}) can be rewritten as
$M^{\mu\nu}_{ab} \bar{A}_\nu^b \bar{A}_\mu^a,$
as well as $\partial_\rho G^{\mu\rho}_a \bar{A}_\mu^a$
plus a total derivative term. Thus, the partition function turns out
to be

\begin{equation}
Z=\sqrt{\pi} \int {\cal D}G \sqrt{det M^{-1}} \int {\cal D} \bar{A}
exp \{ -\int[\alpha G_{\mu\nu}^a G^{\mu\nu}_a - G^{\mu\nu}_a
F_{\mu\nu}^a(\bar{A})]dx \} \delta(2 \bar{A}_\mu^a+2 {M^{-1}}_{\nu\mu}^{ab}
\partial_\rho G^{\rho\nu}_b),
\end{equation}
where the factor 2 in the delta function was introduced for convenience.
If now in this expression the square root of the determinant and the Dirac
delta function are written as exponentials, after a partial integration we
get
\begin{equation}
Z=\sqrt{\pi} \int {\cal D}G {\cal D} \bar{A}{\cal D}\Omega{\cal D}\Lambda
exp \big( \int \{G^{\mu\nu}_a [-\alpha G_{\mu\nu}^a + F_{\mu\nu}^a (\bar{A})-
2{\cal D}_\mu^{(\bar{A})}\Lambda_\nu^a]- {M^{-1}}^{\mu\nu}_{ab}
\Omega_\mu^a\Omega_\nu^b\}dx \big) ,
\end{equation}
which after some manipulations turns out to be
\begin{equation}
Z=\pi \int {\cal D}G {\cal D} \tilde{A}
exp \bigg( \int G^{\mu\nu}_a [-\alpha G_{\mu\nu}^a + F_{\mu\nu}^a
(\tilde{A})]dx \bigg),
\end{equation}
where $\tilde{A}=\bar{A}-\Lambda.$ This result shows the way back to the
model we started with, as well the covariance of the partition function
(\ref{z1}) \cite{Ganor}.

Now we follow the procedure stated above, to find (\ref{z1}) for the
non-Abelian
case with a $CP$-violating $\theta$-term,
 on the manifold $X$ described as above. The
action can be written as

\be
 I_{YM, \theta} = { 1\over 8 \pi} \int_X d^4 x  \bigg( { 4\pi
\over g_{YM}^2} {\rm tr} [ F_{\mu\nu}F^{\mu\nu}] + {i \theta \over 4
\pi}\epsilon_{\mu\nu\rho\sigma} {\rm tr} [F^{\mu\nu} F^{\rho\sigma}] \bigg),
\label{3}
\ee
 $\theta$
is the non-Abelian theta-vacuum and $g_{YM}$ is the Yang-Mills coupling
constant. Equivalently,
\be
 I_{YM, \theta} = { i \over 8 \pi} \int_X d^4 x \bigg( \bar{\tau}
{\rm tr} [F^+_{\mu\nu} F^{+\mu\nu}] - \tau {\rm tr} [F^-_{\mu\nu}F^{-\mu\nu}] \bigg),
\label{4}
\ee
here $ \tau = {\theta \over 2 \pi} + { 4 \pi i \over g_{YM}^2}$ and
$\bar{\tau}$ its complex conjugate,  $F = dA - A\wedge A$, and
$F^{\pm}_{mn}$ are the self-dual and anti-self-dual field strengths
respectively.  We will obtain a Lagrangian which seems to be the corresponding
S-dual Lagrangian to (\ref{3}) employing, as before,  an appropriate
version of the Ro$\check{c}$ek-Verlinde procedure
\cite{Rocek}.   As already shown,  one
finds  the dual Lagrangian of (\ref{4})
\cite{Lozano,Ganor,Mohammedi} to be of the form

\be
\begin{array}{ll}
\tilde{I}_{YM,\theta} & = { i \over 8 \pi} \int_X d^4 x \bigg( -
{1\over \bar{\tau}} G^{+a}_{\mu\nu} G^{+\mu\nu}_{~~~~a} + {1
\over \tau} G^{-a}_{\mu\nu} G^{-\mu\nu}_{~~~~a} + 2
(M^+)^{-1ab}_{\nu\mu} \partial_\rho G^{+\rho \mu}_{~~~~a}
\partial_\sigma  G^{+\nu\sigma}_{~~~~b} \\
& -2 (M^-)^{-1 ab}_{\nu\mu} \partial_\rho
G^{-\rho \mu}_{a} \partial_\sigma G^{-\nu \sigma}_b \bigg),
\label{6}
\end{array}
\ee
where $G^\pm$ are, as mentioned, arbitrary two-forms on $X$ and $M^\pm$ are,
as previously, the adjoints of $G^\pm$, correspondingly.

In this paper we attempt to explore a possible analog to these kind of
``$S$-dual" theories for the gravitational sector. We will show that these
types of theories in fact exist. We will consider only the non-dynamical
gravitational sector as a first step to gain some insight about the much
more involved dynamical sector.

The paper is organized as follows. In Sec. 2 we propose an intermediate
gravitational Lagrangian which interpolates between  our original
topological gravitational Lagrangian and its dual. This dual gravitational
Lagrangian is computed following the procedure given in
\cite{Ganor,Mohammedi}. Sec. 3 is devoted to make a similar procedure but
now including  a $BF$ non-abelian gravitational Lagrangian. Sec. 4 is
devoted to final remarks.

\vskip 2truecm

\section{The Gravitational Analog}

We will first show our procedure to define a gravitational ``$S$-dual"
Lagrangian,  by beginning with the
non-dynamical topological gravitational action of the general form (1.1)

\be
I^{TOP} =  {\Theta^E_G \over 2 \pi} \int_X {\rm tr} R \wedge \tilde R
 + {\Theta^P_G \over 2 \pi} \int_X {\rm tr} R\wedge R,
\label{uno}
\ee
$X$ is a four dimensional closed lorentzian manifold, {\it i.e}
compact, without boundary $\partial X = 0$ and with lorentzian signature.
In this action, the coefficients are the gravitational analogues of
the $\theta$-vacuum in QCD \cite{Deser,Ashtekar}.  Another actions including
gravitational $\Theta$-terms have been analyzed recently in \cite{Smolin}.

This action can be written in terms of the self-dual and anti-self-dual
parts of the Riemann tensor as follows

\be
 I^{TOP} = \int_X \bigg( \tau^+ {\rm tr} R^+ \wedge  R^+ - \tau^-
{\rm tr} R^- \wedge R^- \bigg),
\label{dos}
\ee
whith $\tau^{\pm} = {1 \over 2\pi} \big( \Theta^E_G \mp  \Theta^P_G
\big).$ In local coordinates on $X$, this action is written as

\begin{equation}
 I^{TOP} = \int_X d^4x  \epsilon^{\mu \nu \rho \sigma} \bigg(
\tau^+ R_{\mu \nu}^{+ab} R^+_{\rho \sigma ab} - \tau^- R_{\mu \nu}^{-ab}
R^-_{\rho \sigma ab} \bigg),
\label{tres}
\end{equation}
where ${{R^{\pm}}_{\mu\nu}}^{ab}={1\over 2}({R_{\mu\nu}}^{ab}\mp {i\over
2} {\epsilon^{ab}}_{cd} {R_{\mu\nu}}^{cd})$  and satisfies

\begin{equation}
{\epsilon^{ab}}_{cd} {{R^\pm}_{\mu\nu}}^{cd}= \pm 2i
{{R^\pm}_{\mu\nu}}^{ab}.
\label{cuatro}
\end{equation}

Self-dual (anti-self-dual) Riemann tensors are defined as well in terms of
the self-dual (anti-self-dual) component of the spin connection
$\omega_\mu^{\pm ab} := {1 \over 2} (\omega_\mu^{ ab} \mp {i \over 2}
\epsilon^{ab}_{\ \ cd} \omega_\mu^{cd})$ as

\begin{equation}
{{R^{\pm}}_{\mu\nu}}^{ab}  = \partial_\mu
{{\omega^\pm}_\nu}^{ab}-\partial_\nu {{\omega^\pm}_\mu}^{ab} + {1 \over 2}
f^{[ab]}_{[cd][ef]} {\omega^\pm}_\mu^{cd} {\omega^\pm}_{\nu}^{ef},
\label{cuatro1}
\end{equation}
with
\begin{equation}
f^{[ab]}_{[cd][ef]} = {1 \over 2} \bigg( \eta_{ce} \delta^a_d
\delta^b_f -     \eta_{cf} \delta^a_d
\delta^b_e + \eta_{df} \delta^a_c
\delta^b_e - \eta_{de} \delta^a_c
\delta^b_f - (a \leftrightarrow b) \bigg).
\label{cinco}
\end{equation}
The Lagrangian (\ref{tres}) can be written obviously  as
${\cal L} = {\cal L}_+ + {\cal L}_-$, where  ${\cal L}_{\pm}=
\epsilon^{\mu\nu\rho\sigma}( \tau^{\pm} {{R^{\pm}}_{\mu\nu}}^{ab}
{R^{\pm}}_{\rho\sigma ab}).$

The Euclidean partition function is defined as \cite{Abe}

\begin{equation}
Z(\tau) = Z(\tau^+,\tau^-)= \int {\cal D}\omega^+{\cal
D}\omega^- exp \Bigg( - \int_X( {\cal L}_+ + {\cal L}_-) \Bigg),
\label{siete}
\end{equation}
which satisfies a factorization $ Z(\tau^+,\tau^-) =
Z_+(\tau^+) Z_-(\tau^-)$ where

\begin{equation}
Z_{\pm}(\tau^{\pm}) = \int {\cal D}\omega^{\pm} exp
\Bigg( - \int_X {\cal L}_{\pm} \Bigg).
\label{ocho}
\end{equation}
It is an easy matter to see from action (\ref{tres}) that the partition
function (\ref{siete}) is invariant under combined shifts of $\Theta^E_G$
and $\Theta^P_G$ for $ \tau^{\pm} \to \tau^{\pm} \mp 2,$ in the case of
spin manifolds  and for non-spin manifolds $\tau^{\pm} \to \tau^{\pm}
\mp 5.$

In order to find a S-dual theory to (\ref{tres})  we follow references
\cite{Ganor,Mohammedi}.   The procedures are similar to those presented
in the Introduction for Yang-Mills theories.
We begin by proposing an intermediate Lagrangian

\begin{equation}
\begin{array}{ll}
L = L_+ +L_- = \\
=  \epsilon^{\mu\nu\rho\sigma}\Bigg[ \tau^+ {{R^+}_{\mu\nu}}^{ab}
{R^+}_{\rho\sigma ab} + i G^{+ ab}_{\mu \nu} \bigg( R^+_{\rho
\sigma ab} - (\partial_{\rho} \omega^+_{\sigma ab} - \partial_{\sigma}
\omega^+_{ \rho ab} + [\omega^+_{\rho}, \omega^+_{\sigma}]_{ab}) \bigg)
\Bigg]  \\
 +  \epsilon^{\mu\nu\rho\sigma}\Bigg[
\tau^- {{R^-}_{\mu\nu}}^{ab} {R^-}_{\rho\sigma ab} + i
G^{- ab}_{\mu \nu} \bigg( R^-_{\rho \sigma ab} -
(\partial_{\rho}
\omega^-_{\sigma ab} - \partial_{\sigma} \omega^-_{ \rho ab} +
[\omega^-_{\rho}, \omega^-_{\sigma}]_{ab}) \bigg) \Bigg],
\end{array}
\label{veintitres}
\end{equation}
where $R^+$ is of course treated as a field  independent on $\omega^+$.

For simplicity,  we focus on the partition function of the self-dual part of
this Lagrangian, for the anti-self-dual part $L_-$ one
can follow the same procedure, the corresponding partition function is

\begin{equation}
Z^*_+(\tau^+) = \int {\cal D} R^+ {\cal D}\omega^+ {\cal D} G^+
exp \Bigg( - \int_X L_+ \Bigg).
\label{veinticinco}
\end{equation}

First of all we would like to regain the original Lagrangian
(\ref{tres}).  Thus, we should integrate over the Lagrange multiplier
$G^+$

\begin{equation}
exp\Bigg(-\int_X L^{**}_+ \Bigg) = \int {\cal D} G^+ exp \Bigg( -
\int_X L_+ \Bigg).
\label{veintiseis}
\end{equation}
The relevant part of the integral is
\begin{equation}
\int {\cal D} G^+ exp \Bigg \{ - i \int_X \epsilon^{\mu \nu \rho
\sigma}
G^{+ab}_{\mu \nu}\bigg( R^+_{\rho \sigma ab} - (\partial_{\rho}
\omega^+_{\sigma ab} - \partial_{\sigma} \omega^+_{ \rho ab} +
[\omega^+_{\rho}, \omega^+_{\sigma}]_{ab}) \bigg) \Bigg \}.
\label{veintisiete}
\end{equation}
Using the formula $ \delta(x) = \int {de\over 2 \pi} exp(iex)$ we find

\begin{equation}
Z^*_+(\tau^+) = -2 \pi \int {\cal D} R^+ {\cal D}\omega^+ \delta
[\epsilon (R^{\pm} - \dots ) ] exp ( - \int_X  \epsilon^{\mu \nu \rho
\sigma}
\tau^+R^{+ab}_{\mu \nu} R^+_{\rho \sigma ab}  ).
\label{veintiocho}
\end{equation}
Thus we get the full original Lagrangian $L^{**} = L^{**}_+ + L^{**}_-=
\epsilon^{\mu \nu \rho \sigma} \big( \tau^+ R^{+ab}_{\mu \nu} R^+_{\rho
\sigma ab} + \tau^- R^{-ab}_{\mu \nu} R^-_{\rho \sigma ab}\big)$, where now
$R^{\pm ab}_{\mu \nu}$ depend on $\omega^{\pm ab}_\mu$  as
usual.

Obviously, as before,  factorization holds, $Z^*(\tau^+,\tau^-) =
Z^*_+(\tau^+) Z^*_-(\tau^-)$.

Now we would like to find the dual Lagrangian $\tilde{L}^{**}$. To do this
we have first to integrate out the variables $R^+$ and $\omega^+$, thus

\begin{equation}
exp \Bigg( - \int_X \tilde{L}_+^{**} \Bigg) = \int {\cal D}R^+ {\cal D}
\omega^+ exp \Bigg( - \int_X L_+ \Bigg).
\label{veintinueve}
\end{equation}
We consider first the integration over $R^+$

\begin{equation}
exp\bigg( - \int_X \tilde{L}^{**}_{\omega +}\bigg) = \int {\cal D} R^+ exp
\Bigg(
- \int_X  \epsilon^{\mu \nu \rho
\sigma}\big(\tau^+ R^{+ab}_{\mu \nu} R^+_{ab\rho \sigma} + i R_{\mu
\nu}^{+ \ \ ab} G^+_{\rho \sigma ab}\big) \Bigg).
\label{treinta}
\end{equation}
The functional integral over $R^+$ is of the Gaussian type and defines the
following Lagrangian

\begin{equation}
L_+ =  \epsilon^{\mu\nu\rho\sigma} \bigg[ \frac{1}{4\tau^+} G^{+~~ab}_{\mu\nu}
G^+_{\rho\sigma ab} -i G^{+~~ab}_{\mu\nu} R^+_{\rho\sigma ab}(\omega^+) \bigg],
\end{equation}
where $R^+_{\rho\sigma ab} (\omega^+)$ is given by (\ref{cuatro1}).  Therefore,
another intermediate Lagrangian emerges,  which otherways would result from
adding the degrees of freedom $G^{\pm ab}_{\mu\nu}$.

Now we integrate in the variable $\omega^+$; using  the fact $
\partial X = 0$ and after some manipulations, the relevant part of the
above Lagrangian necessary for the integration in $\omega^+$ is

\begin{equation}
\tilde{L}^{**}_{\omega +} = \dots - 2i \epsilon^{\mu \nu \rho \sigma}
\omega_{\mu}^{+ab} \partial_{\nu} G^+_{\rho \sigma ab} + {i \over
2}\epsilon^{\mu \nu \rho \sigma}  f^{[ab]}_{[cd][ef]}
\omega_{\mu}^{+cd} \omega_{\nu}^{+ef} G^+_{\rho \sigma ab} + \dots
\ \ \ .
\label{tdos}
\end{equation}
Before performing the functional integration it is convenient
to define

\begin{equation}
{M^+}^{\rho \sigma}_{cdef} \equiv i \epsilon^{\mu \nu \rho
\sigma} f^{[ab]}_{[cd][ef]} G^+_{\mu \nu ab}.
\label{ttres}
\end{equation}
Inserting (\ref{tdos}) into (\ref{veintinueve}) and integrating out with
respect to $\omega^+$, we finally find

\begin{equation}
\tilde{L}_+^{**} = +2 \epsilon^{\mu \nu \rho \sigma}
\partial_{\nu} G^+_{\rho \sigma ab} \big(M^+ \big)^{-1 \ abcd}_{\mu
\lambda}
\epsilon^{\lambda \theta \alpha \beta}
\partial_{\theta} G^+_{\alpha
\beta cd} + \dots \ \ \ .
\label{tcuatro}
\end{equation}

Therefore, the complete dual Lagrangian is

\begin{equation}
\begin{array}{ll}
\tilde{L}^{**} & =  \tilde{L}_+^{**} + \tilde{L}_-^{**} \\
&= \epsilon^{\mu\nu\rho\sigma} \big[ -{1 \over 4
\tau^+} G^{+ab}_{\mu \nu} G^+_{\rho\sigma ab} + {1 \over 4
\tau^-} G^{-ab}_{\mu \nu} G^-_{\rho \sigma ab} +
2 \partial_{\nu} G^+_{\rho \sigma ab} {(M^+)}^{-1 \ abcd}_{\mu
\lambda} \epsilon^{\lambda \theta \alpha \beta}
\partial_{\theta} G^+_{\alpha \beta cd} \\
&- 2 \partial_{\nu} G^-_{\rho
\sigma
ab} {(M^-)}^{-1 \ abcd}_{\mu
\lambda} \epsilon^{\lambda \theta \alpha \beta}
\partial_{\theta} G^-_{\alpha \beta cd}\big].
\end{array}
\label{tcinco}
\end{equation}

Of course, the condition of factorization for the above Lagrangian still
holds.

\vskip 2truecm

\section{INCLUDING A $BF$ GRAVITATIONAL TERM}

In the previous section we found a dual Lagrangian for a gravitational
topological model. Therefore, we have worked out therefore a non-dynamical
gravitational system. We would like to extend these results to a theory
which contains dynamical gravity. This problem is not
easy, so we shall  try first a toy model. In this section we
consider a four-dimensional non-abelian $BF$-theory \cite{Gary,Baez},
coupled to our topological field theory $L_{TOP}$

\begin{equation}
L = L_{BF} + L_{TOP},
\label{luno}
\end{equation}
where $L_{BF}= \alpha {\rm tr} ( \Sigma \wedge R) + \beta {\rm tr} ( \Sigma
\wedge
\Sigma)$ and $ L_{TOP} = \gamma {\rm tr} ( R \wedge \tilde{R}) + \delta
{\rm tr} ( R
\wedge R)$ with $\alpha$ and $\beta$ parameters of the $BF$ theory, and
$\gamma$ and $\delta$  proportional to the gravitational $\Theta$ angles,
defined  in the previous section. Thus,  we can see this Lagrangian as the
non-abelian $BF$-theory to which we add a topological (non-
dynamical)  gravitational Lagrangian, in a similar spirit as we dealt with
the Yang-Mills case in Eq. (\ref{2}).

Now, proceeding as in the last section,  we consider only the
self-dual terms

\begin{equation}
{\cal L}_+=  \epsilon^{\mu \nu  \rho \sigma} \bigg(
A R^{+ab}_{\mu \nu}R^{+}_{\rho \sigma a b} +
B \Sigma^{+ab}_{\mu \nu} R^{+}_{\rho \sigma a b} + C \Sigma^{+ab}_{\mu \nu}
\Sigma^{+}_{\rho \sigma a b}   \bigg),
\label{ldos}
\end{equation}
where $A, B$ and $C$ are appropriate constants.  This action  looks
similar to the
Pleba\'nski-Ashtekar dynamical action \cite{Pleban,Samuel}.  Hence, by
studying the action (\ref{ldos}),  we hope to learn how to deal with the
dynamical case.

In order to dualize ${\cal L}_+$,  let us propose, as before, the
{\it intermediate
self-dual Lagrangian}

\begin{equation}
\begin{array}{ll}
L_+= \epsilon^{\mu \nu \rho \sigma} \bigg(A R^{+ ab}_{\mu \nu}
R^{+}_{\rho \sigma ab}
+ B \Sigma^{+ab}_{\mu \nu} R^{+}_{\rho \sigma ab} + C \Sigma^{+ab}_{\mu \nu}
\Sigma^{+}_{\rho \sigma ab} + E G^{+ab}_{\mu \nu} \\
\big( R^{+}_{\rho \sigma ab} -
\partial_{\rho} \omega^{+}_{\sigma ab}
+ \partial_{\sigma}
\omega^{+}_{\rho ab} + \frac{1}{2} f_{[ab][cd][ef]} \omega^{+ cd}_{\rho}
\omega_{\sigma}^{+ef} ) \bigg),
\end{array}
\label{ltres}
\end{equation}
where $G^{+ab}_{\mu \nu}$ is a Lagrange multiplier and $E$ is a
constant.
The self-dual sector of the Euclidean partition function is

\begin{equation}
Z^*_+ = \int {\cal D} R^+ {\cal D}\omega^+ {\cal D}\Sigma^+ {\cal D} G^+
 exp \Bigg( - \int_X L_+ \Bigg),
\label{lcuatro}
\end{equation}
where $R^+$ is of course treated as an independent field.

Following the same procedure as in section 2, we will compute
$\tilde{L}^{**}_+$,  the
dual   of the Lagrangian (\ref{ldos}). It is
defined by

\begin{equation}
\begin{array}{ll}
exp \Bigg( - \int_X \tilde{L}_+^{**} \Bigg) = & \int {\cal
D} \omega^+{\cal D}\Sigma^+ {\cal D} R^+ exp \Bigg( - \int_X L_+ \Bigg) \\
& = \int {\cal
D} \omega^+{\cal D}\Sigma^+  exp \Bigg( - \int_X L_{\Sigma+} \Bigg) \\
& =\int {\cal
D} \omega^+ exp \Bigg( - \int_X L_{\omega+} \Bigg). \\
\end{array}
\label{lcinco}
\end{equation}
The first integration respect to $R^+$ gives

\begin{equation}
\begin{array}{ll}
 L_{\Sigma +} & =  \epsilon^{\mu\nu\rho\sigma} \bigg[ - \frac{1}{4A} (B
\Sigma^{+ab}_{\mu \nu} + E G^{+ab}_{\mu \nu})(B
\Sigma^{+}_{\rho \sigma ab} + E G^{+}_{\rho \sigma ab}) \\
& + C \Sigma^{+ ab}_{\mu\nu} \Sigma^+_{\rho\sigma ab}+
E G^{+ ab}_{\mu\nu} (\partial_\rho \omega^+_{\sigma ab} -
\partial_\sigma \omega^+_{\rho ab} + {1\over2} f_{[ab]}^{[cd][ef]}
\omega_{\rho cd} \omega_{\sigma ef} \bigg) \bigg]. \\
\end{array}
\label{lseis}
\end{equation}
The relevant part of this Lagrangian for the integration in $\Sigma^+$ is

\begin{equation}
L_{\Sigma+} = \epsilon^{\mu \nu \rho \sigma} \bigg[\dots + \bigg(C -{B^2
\over 4A}\bigg) \Sigma^{+ab}_{\mu \nu} \Sigma^{+}_{\rho \sigma ab} -
{BE\over 2A} G^{+ab}_{\mu \nu}\Sigma^{+}_{\rho \sigma ab} +\dots \bigg],
\label{lsiete}
\end{equation}
which is of the Gaussian type. Integration gives

\begin{eqnarray}
L_{\omega +} &=&  {E^2\over 4A} \bigg( {B^2 \over \Delta} -1 \bigg)
\epsilon^{\mu\nu \rho \sigma} G^{+ab}_{\mu \nu} G^{+}_{\rho \sigma ab} + 2 E
\epsilon^{\mu\nu \rho \sigma} \partial_\rho G^{+ab}_{\mu \nu}
\omega^{+}_{\sigma ab}  \nonumber \\
&+& {1\over2} E \epsilon^{\mu \nu \rho \sigma} f_{[ab][cd][ef]}  G^{+ab}_{\mu
\nu} \omega^{+ cd}_{\rho}\omega^{+ ef}_{\sigma},
\label{locho}
\end{eqnarray}
where $\Delta = B^2 -4AC$.  Finally, the computation of $\int {\cal D}\omega^+
exp\Bigg(- \int_X L_{\omega +} \Bigg)$ gives

\begin{equation}
\tilde{L}^{**}_+ =  {E^2\over 4A} \bigg( {B^2 \over \Delta} -1 \bigg)
\epsilon^{\mu\nu \rho \sigma} G^{+ab}_{\mu \nu} G^{+}_{\rho \sigma ab} -
2E \epsilon^{\mu\nu \rho \sigma} \partial_\nu G^{+ab}_{\rho \sigma}
(M^{+})^{-1}_{\mu \lambda abcd} \epsilon^{\lambda \theta \rho \sigma}
\partial_{\theta} G^{+~~cd}_{\rho \sigma},
\label{lnueve}
\end{equation}
where $M^{+\rho\sigma}_{ab~cd} \equiv \epsilon^{\mu\nu\rho\sigma}
f_{[ab][cd]}^{~~~~~[ef]} G^+_{\mu\nu ef}$.

The full action can be finally constructed from
\begin{equation}
\tilde{L}^{**} = \tilde{L}^{**}_+ + \tilde{L}^{**}_-.
\label{ldiez}
\end{equation}
Obviously,  Lagrangian (\ref{ldiez}) contains terms of the form
$\Sigma \widetilde{\Sigma}$
and $R \widetilde{\Sigma}$ that will add to (\ref{luno}),  giving then a generalized
$BF$-theory.
The coefficient of the $G^2$-term in (\ref{lnueve}) can be adjusted to be
the inverse of $A$ or $B$ in (\ref{ldos}). We can
observe that in this case the dual transformation involves a more
complicated law of transformation for the coupling constants.

\section{\bf Final Remarks}
\setcounter{equation}{0}

In this paper we have defined a gravitational analog of $S$-duality
following similar procedures to those well known in non-abelian
non-supersymmetric Yang-Mills theories \cite{Ganor,Mohammedi}.

We have shown, how  to construct `dual' Lagrangians
for  pure topological gravity and
for a  $BF$ theory coupled to  topological gravity. In the
computation of `dual' Lagrangians we have not considered fermionic
determinants of zero modes \cite{Abe}. We expect that the computation of
these determinants will give the manner of transforming the
partition function at the quantum level. This should give rise to some analog
of
`modular weights' \cite{Wittenone}.

We are aware that this analogy only was carried out at the level of the
structure group of the frame bundle over $X$ and not over all the genuine
symmetries which arise in Einstein gravity theories, such as Diff$(X)$.
Related to this, it is interesting to note that in our `dual' Lagrangians
in Eqs. (\ref{tcinco}), and
(\ref{ldiez}) only  partial derivatives of the $G's$ fields appear instead of
covariant derivatives. However, it can be shown that following a procedure
similar to that presented in the introduction, one can get the covariance
of the partition functions. As it has been pointed out by Atiyah
\cite{Atiyah}, $S$-duality symmetry in field theory is a duality
between the fundamental homotopy group of the circle and the space of
group characters of representation theory of the circle. That means a sort
of `identification' between algebraic properties of a topological space.
It would be very interesting to investigate what is the mathematical
interpretation of the `gravitational $S$-duality'. The way how this
`gravitational $S$-duality' can be mixed with the usual $S$-duality in
field theory and string theory, are now under current
investigation. Finally, in a forthcoming paper we will exhibit a ``dual"
theory to   dynamical gravity. For this
purpose, the MacDowell-Mansouri
gauge theory of gravity \cite{MM} is worked out.

\vskip 2truecm

\centerline{Acknowledgements}

This work was supported in part by CONACyT grants 3898P-E9608 and
4434-E9406.  One of us (H.G.-C.) would like to thank CONACyT for support
under the program {\it Programa de Posdoctorantes: Estancias Posdoctorales
en el Extranjero para Graduados en Instituciones Nacionales 1996-1997}, and
Academia Mexicana de Ciencias (AMC) for partial support  under the
program
{\it Estancias de Verano para Investigadores J\'ovenes}.

\end{document}